\RequirePackage{fix-cm}
\documentclass[smallextended]{svjour3}       
\smartqed  
\usepackage{graphicx}
\usepackage{array}
\usepackage{mathptmx}      
\usepackage{amssymb}
\usepackage[usenames,dvipsnames]{xcolor}
\definecolor{bros}{rgb}{0.85,.66,0}
\begin{document}

\title{Unifying Geometrical Representations of Gauge Theory}


\author{Scott Alsid         \and
        Mario Serna 
}


\institute{S. Alsid and M. Serna \at
              2354 Fairchild Drive, Department of Physics\\ United States Air Force Academy, CO  80840 \\
              Tel.: +1-719-333-3510\\
              Fax: +1-719-333-3182\\
              \email{c15scott.alsid@usafa.edu}
              \email{mario.serna@usafa.edu}
}

\date{Received: date / Accepted: date}

\maketitle

\begin{abstract}
We unify three approaches within the vast body of gauge-theory research that have independently developed distinct representations of a geometrical surface-like structure underlying the vector-potential.   The three approaches that we unify are: those who use the compactified dimensions of Kaluza-Klein theory, those who use Grassmannian  models (also called gauge theory embedding or $CP^{N-1}$ models) to represent gauge fields, and those who use a hidden spatial metric to replace the gauge fields.
In this paper we identify a correspondence  between the geometrical representations of the three schools.
Each school was mostly independently developed, does not compete with other schools, and attempts to isolate the gauge-invariant geometrical surface-like structures that are responsible for the resulting physics.  By providing a mapping between geometrical representations, we hope physicists can now isolate representation-dependent physics from gauge-invariant physical results and share results between each school.  We provide visual examples of the geometrical relationships between each school for $U(1)$ electric and magnetic fields.  We highlight a first new result: in all three representations a static electric field (electric field from a fixed ring of charge or a sphere of charge) has a hidden gauge-invariant time dependent surface that is underlying the vector potential.
\end{abstract}

\section{Introduction}
\label{intro}

In this study, we unify three small but largely independently developed schools
within the vast body of gauge-theory research that have developed distinct representations of a geometrical surface-like structure underlying the vector-potential.
By school we mean a grouping of conceptual approaches which share a common methodology.  The approaches are not in competition with each other.  They are simply our grouping of mathematical tools that make use of a surface-like representation from which one can derive or induce a gauge field.
Each school has been employed by Fields Medalist and Nobel Prize winners to extract {or to separate} gauge-invariant key physics from gauge-dependent artifacts.  Each school has been largely independently invented; each school has had distinct strands of papers with very little reference to papers of other schools.  We highlight the easily overlooked commonalities of the different strands within each school, and then we tie the geometric representations of each school onto a common representation.  This paper's new results are: the direct geometrical relationship between each school, the explicit examples that we work out, and third we will show that in all three representations, a static electric field has a hidden time dependence that is not captured by our normal notation.

\begin{figure}[b]
\centerline{
\includegraphics[width=1.0\textwidth]{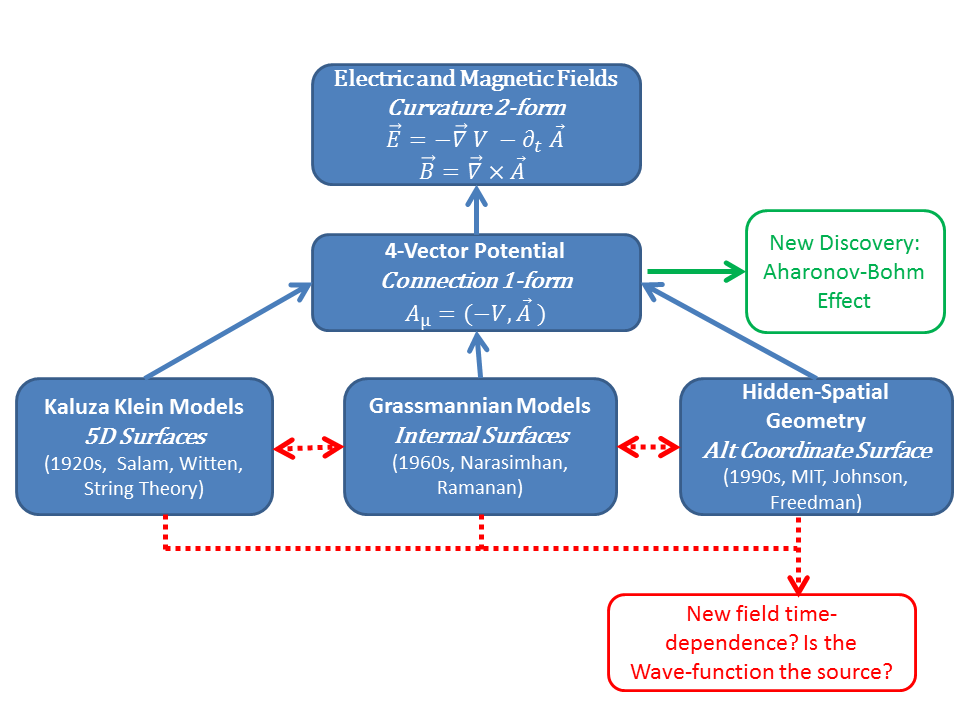}}
 \caption{\label{FigMapGaugeTheory} Map of geometric foundations of gauge theory.  This paper contributes the connections depicted in the dotted lines.}
\end{figure}

Although the `spell' of gauge theory has captured most modern physicists, most of the research on gauge theory does not fall into one of the three schools that we describe.
Fig.~\ref{FigMapGaugeTheory} shows a map of gauge theory and where this paper contributes. Our contribution, as depicted in Fig.~\ref{FigMapGaugeTheory}, is represented by the dotted red lines.

  Historically electric and magnetic fields were thought to be the fundamental objects in the model.
This is the top layer in the figure.  Vector potentials were introduced as a mathematical trick, but were not ascribed as physical objects in the model.  It was not until the  the Aharonov-Bohm effect was predicted and observed that the vector potential was elevated from a representational convenience to something with predictive power.
As depicted in the Fig.~\ref{FigMapGaugeTheory} vector potentials are a layer deeper as we dig for the foundations of gauge theory.
Today we recognize that electromagnetic fields are  a curvature $2$-form that originates from the vector potential, which is a connection $1$-form.

But what is the geometrical surface that gives rise to this $1$-form?
This is the deeper foundation of gauge theory that is represented on the third row down on Fig.~\ref{FigMapGaugeTheory}.
There have been at least three schools providing possible geometrical surface-like foundations to the vector potential. This paper reviews these three schools and explores the relationships between them depicted by the red dotted lines.  In this paper we will observe that in all three schools there is a gauge-invariant hidden time dependence to the surface-like geometrical structures of electric fields that is not captured in the connection $1$-form representation nor in the curvature $2$-form representation.

Before going into details on the three representations, one might ask why different representations might be expected to give new physics? Richard Feynman once remarked: ``every theoretical physicist who is any good knows six or
seven different theoretical representations for exactly the same
physics. He knows that they are all equivalent $\ldots$ but he
keeps them in his head hoping that they will give him different
ideas for guessing (new physical laws)" \cite[pg 168]{feynman1994character}.
Therefore, we do not expect new physics at this stage.  We do hope that finding the commonalities between deeper representations of gauge theory will provide insights that may help us ``guess'' new physical laws. The time-dependence of electric fields in the surface-like layer is one such insight.

The first attempt to find a surfrace underlying the vector potential was started by Theodor Kaluza and Oskar Klein \cite{Kaluza:1921ar,Klein:1926ar}. They use a $4+n$ dimensional space-time where extra dimensions are curled up and result in a gauge theory
(see \cite{Schwarz:1992zt,Salam:1981xd} for reviews).  This well-known school has over 1600 papers. The second school uses a Grassmannian manifold to represent gauge fields using a type of gauge-theory embedding
\cite{Narasimhan1961,Narasimhan63,79Atiyah,Corrigan:1978ce,Dubois-Violette:1979it,Felsager:1979fq,Cahill:1993mp,Cahill:1993uq,Cahill:1996yw,Valtancoli:2001gx,Bars:1978xy,Bars:1979qd,Stoll:1994cn,Stoll:1994vx,PhysRevLett.52.2111,Serna:2002ux,Serna:2005ar,Gliozzi:1978xe,Eichenherr:1978qa,Gava:1979sp,1980NuPhB.174..397D,Balakrishna:1993ja,Palumbo:1993vu,PhysRevD.66.025022}
\cite{Marsh:2007qp,2006JPhA...39.9187G,2010JMP....51j3509H}.
The third school introduces alternative variables for gauge theory that uncover a hidden spatial metric which reproduces the gauge fields
\cite{Goldstone:1978he,Freedman:1993mu,Freedman:1994rg,Lunev:1994ty,Haagensen:1994sy,Haagensen:1995py,Schiappa:1997yh,Niemi:2010mw,Zee:1988mc}.
Each of the schools start with a different geometrical representation which then faithfully maps onto the traditional gauge fields $A_\mu$.
This paper directly unifies the geometrical representations of the three schools without appealing to their common gauge-field image-space.
Our unified geometrical representation allows physicists to better identify gauge-invariant foundations underlying the physical results.
As an example, we will discuss a hidden time dependence that we reveal is present even in static electric fields.  Our unification will also help translate results, such as instanton solutions, between each independent school.

Mathematicians describe both gauge theory and Riemannian manifolds with the language of fiber bundles.  Fiber bundles are not a geometrical representation, but rather a rigorous lexicon used to describe a wide array of geometrical structure.
This language enables descriptions of gauge theories on topologically non-trivial spaces.
However, the power gained by abstraction to the fiber bundle language often leaves out insight that may be gained from explicit examples.
Here we concern ourself only with local descriptions of gauge fields on topologically trivial flat space-time; therefore, we will avoid extensive use of the bundle language in favor of explicit examples.

This paper is organized as follows:  In section \ref{sec2} we review the development and research activity of each school.  Sections \ref{sec3} and \ref{sec4} contain our new research results: they present the connections between the Kaluza-Klein and Grassmannian school, and between the Hidden-Spatial-Geometry and the Grassmannian school respectively.  Finally in section \ref{Sec5}  we provide examples with familiar electric and magnetic fields.  In our conclusion, we discuss the hidden time dependence that is revealed to be in static electric fields.

\section{Literature Survey of the Three Schools}
\label{sec2}

       \begin{figure}
             \centering
                 (a) \includegraphics[width=0.45\textwidth]{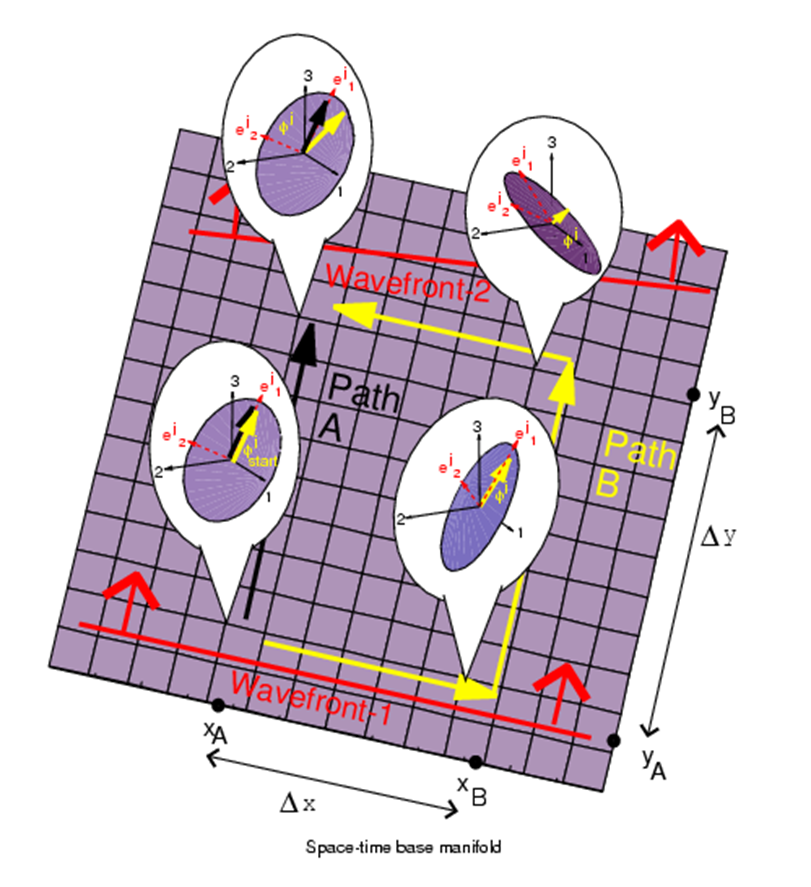}
                 (b) \includegraphics[width=0.4\textwidth]{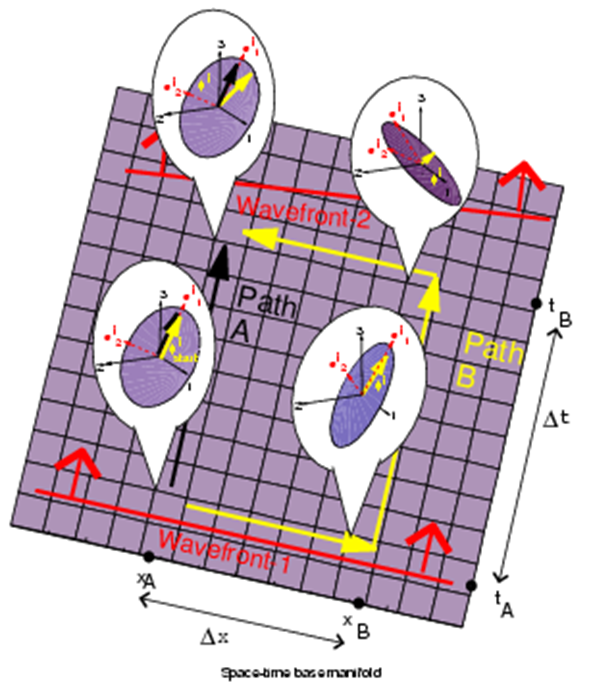}

         \caption{\label{magdisk}\label{elecdisk}Shown is a geometrical representation of the magnetic field (left) and the electric field (right).  The angular change in the phase of a wave function after parallel transport around a closed loop in space-time yields electric and magnetic fields multiplied by the area of the loop enclosed.  This parallels Riemannian geometry where the Riemann tensor gives the rotation matrix that results from parallel transport around a loop.}
         \end{figure}

All the geometrized representations that we discuss emphasize non-integrable phase factors to define the internal curvature \cite{Wu:1975es}.
The Wilson loop $\Delta \theta = \oint{\vec{A} \cdot d\vec{r}}$ gives the phase-angle shift\footnote{We have chosen to work in units where $\hbar = c  = 1$ and where we absorb the electron's charge $e$ into the definition of $A_\mu$.} resulting from parallel transport of the wave function around an infinitesimal loop.  The non-integrable phase is similar to how curvature is found in Riemannian geometry.

For example, the magnetic field is equal to the phase-angle change in the wave function after parallel transporting the wave function around a closed loop on a spatial slice of space-time.
In the limit of an infinitesimal loop, the magnetic field is given in terms of the phase-shift $\Delta \theta$ as
\begin{equation}
\label{Bmag}
B_z = \frac{\Delta \theta}{\Delta x \Delta y} = \frac{\displaystyle \oint{\vec{A} \cdot d\vec{r}}}{\Delta x \Delta y} = \frac{\displaystyle \int \int ({\vec{\nabla} \times \vec{A}}) \cdot (d\vec{x} \times d\vec{y})}{|d\vec{x} \times d\vec{y}|}
\end{equation}
where we have used the Wilson loop and classic vector identities.

Figs.~\ref{magdisk}a and \ref{elecdisk}b show this non-integrable phase angle using the tools of the Grassmannian school described in section \ref{embedcamps}.  In this figure, the complex plane on which the wave function lives is represented by the plane spanned by the two red basis vectors.  The complex plane is inserted into a trivial internal space at each space-time point and is represented by a disk.  A wave function is shown as a vector (black or yellow) on the disk inserted at each space-time point.  We parallel transport the wave function along two paths (A and B) represented by the black and yellow vectors.  Comparing path A with path B gives the non-integrable phase shift $\Delta \theta$.  Because of this phase shift, the wave front of a plane wave is pulled and the plane wave changes directions.

Fig.~\ref{magdisk}a shows the case of the magnetic field where the loop is all spatial.   Likewise, Fig.~\ref{elecdisk}b shows the electric field is equal to parallel transporting the wave function or matter field around a closed loop on a part spatial and part temporal slice of space-time.

As we delve into a review of the three schools, there will be a proliferation of notations for each of the schools and the past papers.  To help clarify this, we provide a table in appendix \ref{AppendixVariableDefs} to help define the different variables as we use it to express the basis vectors in each school and the various index types.

\subsection{The Kaluza-Klein School}
\label{Sec2.1}
Kaluza Klein theories unify classical electromagnetism with Einstein's general relativity \cite{Kaluza:1921ar} \cite{Klein:1926ar}.  They posit extra spatial dimensions that are compactified within ordinary space-time along a very small radius $R$.  All tensor quantities are independent of this fifth coordinate (the cylinder condition).

In traditional Kaluza-Klein theory the line element of the five-dimensional space is\footnote{Throughout this paper we use the convention that lower-case Latin letters near the beginning of the alphabet $a,b,...$ will be gauge-theory color indices, Greek letters $\mu, \nu,...$ will be space-time coordinates, upper-case Latin letters $A,B,...$ will be used for Kaluza-Klein metric indices, and lower-case Latin letters towards the middle of the alphabet $i,j,...$ will be used for the variables corresponding to subspaces of space-time and the embedding dimensions, where context will keep them distinct.  The Kaluza-Klein index values 0 through 3 are the usual space-time coordinates $t,x,y,z$ and the index value 5 is the fifth dimension coordinate $x^5$, which is used to parameterize the tiny compact dimension. The appendix provides a summary.}
\begin{equation}
ds^2 = g_{\mu\nu}dx^\mu dx^\nu + (R\, A_\mu \,dx^\mu + dx^5)^2,
\end{equation}
where we omit the dilaton field for simplicity of presentation.   Here $g_{\mu\nu}$ is the familiar four-dimensional metric from general relativity, $A_\mu$ is the four-vector potential, $x^5$ is the fifth dimension's coordinate, and $R$ is the radius of the curled up fifth dimension.

In Kaluza-Klein theory charge is explained as motion of a neutral particle along the fifth dimension, where the two directions it can go in $x^5$ explain the two different types of charge.  Electric fields are four-dimensional manifestations of the inertial-dragging effect in the fifth dimension \cite{Gron85} \cite{Gron92} \cite{Gron:2005aw}.  Furthermore coordinate transformations of the fifth dimension are shown to be $U(1)$ gauge transformations.

One pitfall of the classical theory is that there are no measurable new predictions.  Another pitfall occurs with quantum mechanics.  The wave function around the fifth dimension gives particles a mass-spectrum tower of $m^2= (n/R)^2$, where $n$ is an arbitrary integer.  For an $R$ near the Planck scale, particles would be either massless or have Planck-scale masses, which implies that the model must be modified to be used in new physical theories.  Modified Kaluza-Klein theories play a large role in string theory.  For a further review of Kaluza-Klein theory see references \cite{Salam:1981xd,Schwarz:1992zt} and the references therein.

\subsection{The Grassmannian School}
\label{Sec2.2}
\label{embedcamps}

Grassmannian representations of gauge fields started in 1961 when Narasimhan and Ramanan showed that every $U(n)$ gauge theory could be represented by a section of a Grassmannian $Gr(n,N)$ fiber bundle  \cite{Narasimhan1961,Narasimhan63}.
A Grassmannian manifold $Gr(n,N)$ is the set of orientations an $n$-plane can take in a larger $N$-dimensional space with a fixed origin.
Another way to view $Gr(n,N)$ fiber bundle is as a $n$-plane embedded into a higher-dimensional $N$-Euclidean space that is inserted into each point in space and time.
The Grassmannian school is essentially a gauge theory version of the 1956 Nash embedding theorem which proved that every Riemannian manifold could be embedded in a higher-dimensional Euclidean space \cite{Nash56}.

In the language of bundles, Narasimhan and Ramanan proved for any $U(n)$ gauge field and $d$ space-time dimensions, the gauge field can be constructed by inserting a $\mathbb{C}^n$ vector bundle into a trivial $\mathbb{C}^N$ vector bundle if $N \geq (d+1)(2d+1)n^3$.  Narasimhan's condition guarantees us an embedding for this $N$, but we can sometimes represent the embedding for specific field configurations for smaller $N$ as we will do in section \ref{Sec5}.  For an $O(n)$ gauge field $\mathbb{R}^n$ vector bundles are embedded in a trivial $\mathbb{R}^N$ vector-bundle.

\begin{figure}[t!]
\centering
\includegraphics[width=\textwidth]{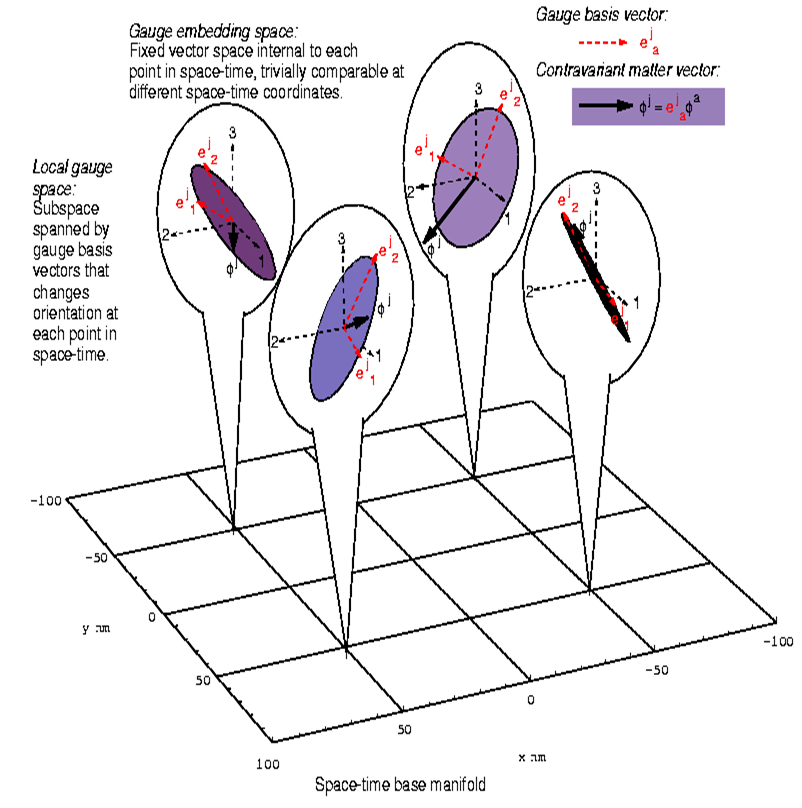}
\caption{A graphical representation of the Grassmannian school. A set of two basis vectors span the internal vector space attached to every point in space-time.  How they vary determines the electromagnetic field.}
\label{Standardpic}
\end{figure}

In the Grassmannian school, wave functions are sections of the $n$-dimensional vector bundle. That is, they are a vector on the $\mathbb{C}^n$ or $\mathbb{R}^n$ vector space.  By definition the vector bundles have a fixed origin.
All the embedding-school approaches have a set of $n$ orthonormal gauge basis vectors $\vec{e}_a$ that span the gauge fiber internal to each space-time point.  The dual basis vectors $\vec{e}^{\;a}$ satisfy $\vec{e}^{\;a} \cdot \vec{e}_b = \delta^a_b$.  There are $n$ vectors $\vec{e}_a$ in a real or complex Euclidean $N$-dimensional embedding space.  The matter field (wave function) exists as a vector on the gauge fiber spanned by the gauge-fiber basis vectors:
\begin{equation}
\vec{\phi} = \phi^a \vec{e}_a.
\end{equation}
The projection operator is the outer product $P^j_k = {e}^j_a {e}^{\;a}_k$.  The gauge field is then
\begin{equation}
\label{A}
(A_{\mu})_{\;b} ^{a}= i\vec{e}^{\;a}\cdot \partial_\mu \vec{e}_b.
\end{equation}
Notice that if $n=N$ then $A_\mu$ is a pure gauge with a vanishing $F_{\mu\nu}$.
In all the cases that we study here, $N>n$.
Fig.~\ref{Standardpic} shows the $Gr(2,3)$ Grassmannian model visually.  The bubbles show the $N=3$ trivial vector space inside each space-time point.  The red vectors are the gauge-fiber basis vectors $\vec{e}_a$ which span the displayed disk. The gauge fields depend on all the ways one can orient the $R^2$ space within the trivial $R^3$ space.  The wave function or matter field $\vec{\phi}$ is the black vector that lives on the disks.

A gauge transformation is a rotation of the basis vectors $\vec{e}_a$ accompanied by the inverse rotation on the matter vector coefficients $\phi^a$ that preserves their inner product and leaves the wave function $\vec{\phi} = \phi^a \vec{e}_a$ and the projection operator $P^j_k = e^j_a {e}^{\;a}_k$ invariant. It is very central to our argument to understand that gauge transformations leave two objects invariant: (1) the plane spanned by the basis vectors $\vec{e}_a$, and (2) the vector formed by the wave function $\vec{\phi}^a$. Global transformations on the embedding space do not affect Eq.(\ref{A}).

A long list of notable physicists have employed the Grassmannian-model as a part of their gauge theory research.
In the following review, we map the notation used in these previous approaches onto the notation introduced above.
Atiyah in 1979 \cite{79Atiyah} defined the linear maps $u_x: \mathbb{R}^n \rightarrow \mathbb{R}^N$, whose image was in the trivial space $\mathbb{R}^N$.  Atiyah's $u$'s play the role of the gauge-fiber basis vectors $\vec{e}_a$.  The projection operator is written as $P = uu^*$, with $u^*u = 1$, and the gauge potential is $A_\mu = u^* \partial_\mu u$, where $u^*$ is the dual to $u$.
  Atiyah, Drineld, Hitchin, and Manin (ADHM) used the rectangular matrices of the Grassmannian school as one of the tools in their construction of self-dual instanton solutions in Euclidean Yang-Mills Theory \cite{Atiyah1978185}.
Corrigan and followers \cite{Corrigan:1978ce,Alekseevsky:2002pi} 
used the embedding representation in finding Green's functions for self-dual gauge fields.
Dubois-Violette \cite{Dubois-Violette:1979it} created a formulation of gauge theory using only globally defined complex $N \times n$ matrices $V$ (analogous to $e^j_{\,a}$) such that $V^\dagger V=I$ and $VV^\dagger=P$, and $A_\mu =V^\dagger (x) \partial_\mu V(x).$

An independent research line refers to the Grassmannian school as $CP^{N-1}$ models
\cite{Eichenherr:1978qa,Gava:1979sp,1980NuPhB.174..397D,Balakrishna:1993ja,Palumbo:1993vu,PhysRevD.66.025022,Marsh:2007qp,2006JPhA...39.9187G,2010JMP....51j3509H}.
In the $CP^{N-1}$  models a setup is created with $z^\dagger \cdot z =1$ where $z$, which is sometimes called a zweibein, is a complex $N$-vector.  The gauge field
$A_\mu = z^\dagger \partial_\mu \cdot z$ is discovered in the equations of motion.  Here the complex vector $z$ plays the role of a gauge basis vector $e^j_a$ with complex dimensions $1 \times N$.

Felsager, Leinaas, and Gliozzi \cite{Gliozzi:1978xe,Felsager:1979fq} had a similar approach.  In a manner very similar to Fig.~\ref{Standardpic} and section \ref{Sec5}, they geometrically represented magnetic fields by use of plane bundles in $\mathbb{R}^3$, where the distribution of the planes in each point was characterized by a curvature related to the magnetic field strength.
For two vectors $\vec{e}_1, \vec{e}_2$ orthonormal to each other and to the normal vector of the plane, the vector potential is $A_j = \lambda \vec{e}_1 \cdot \nabla_j \vec{e}_2,$ where the $\vec{e}$'s play the same role as $\vec{e}_a$ introduced in the beginning of this section, and $\lambda$ is a constant for dimensionality. Since the nineties, Cahill \cite{Cahill:1993mp,Cahill:1996yw,Cahill:1993uq,PhysRevD.88.125014} has used gauge basis vectors $\vec{e}_a$ in lattice simulations and in his most recent textbook \cite[Sections 11.51 and 11.52]{cahill2013physical}.

In finding projectors for the fuzzy sphere, Valtancoli \cite{Valtancoli:2001gx} used the connection $A_n^\nabla = \langle \psi_n, d \psi_n \rangle$ for $n$-monopoles.  Here $|\psi\rangle$ plays the role of $\vec{e}_a$.

In another variant of the Grassmannian school, Bars \cite{Bars:1978xy,Bars:1979qd} used a separate embedding for each of the spatial dimensions of the gauge field (corner variables), as opposed to using a single embedding for all the gauge-fiber basis vectors.  He used $n \times n$ unitary matrices $B_{13}^{ij}, B_{23}^{ij}$ to rewrite the canonical variables $A_i^a$ and $E_i^a$.  For example, $A_1^a$ was written as $T^a A_1^a = i B_{13}^\dagger \partial_1 B_{13},$ where $T^a$ is a generator of $SU(n)$.
Stoll \cite{Stoll:1994vx,Stoll:1994cn} introduced angle variables in the Hamiltonian formulation of QCD to investigate the low-energy properties in terms of gauge invariant degrees of freedom.  The angle variables are similar to corner variables and are the exponents of $SU(n)$ matrices, and the gauge fields are defined as $A_j(x) = \frac{i}{g} V_j(x) \partial_j V_j^\dagger (x) \; {\rm{(no \; summation)}}$.

Zee and Wilczek,  building on work by Simon, also independently developed a Yang-Mills structure associated with Barry's phase and degenerate spaces (see \cite{Simon:1983mh,PhysRevLett.52.2111,Zee:1988mc,Zee:2003mt}).  A given wave function is expanded in terms of eigenfunctions spanning a degenerate subspace  $\Psi(t) =  c_a \psi_a(t)$.  One finds in the adiabatic limit that $\frac{dc_b}{dt} = -A_{ba} c_a$, where $A_{ba}(t) = i\langle \psi_b(t)|\frac{\partial \psi_a}{\partial t}\rangle$.
For a Hamiltonian $H(t)$ that depends on parameters $\lambda^1,...,\lambda^d$, when one traces out a path in the parameter space the time derivative of $c_b$ becomes $\frac{dc_b}{dt} = - (A_\mu)_{ba} c_a \frac{d \lambda^\mu}{dt}$, where $(A_\mu)_{ba} = i\langle \psi_b | \partial_\mu \psi_a \rangle.$  In Zee and Wilczek's approach, $|\psi_a\rangle$ plays the role of $\vec{e}_a$.

In the early 2000's one of us (MS) and Cahill used the Narasimhan and Ramanan theorem to visualize the geometry of simple electromagnetic fields with an $SO(2)$ gauge group. To gain some visual intuition they found $SO(2)$ gauge basis vectors $\vec{e}_a$ embedded in an $\mathbb{R}^3$ trivial fiber for certain vector potentials \cite{Serna:2002ux}.  For a matter vector on the gauge fiber, as represented visually in their work, a clockwise rotation in the momentum direction corresponded to a positive charge, while a counterclockwise rotation corresponded to a negative charge.  In addition to this they observed an indication for a geometry-based explanation of charge quantization.  This is similar to the representation of charge in Kaluza-Klein.
Although all free fundamental particles have $\pm e$ charge, quarks have fractional charge.  The fractional charge would follow from a GUT gauge theory, such as that of $U(1) \times SU(2) \times SU(3) \subset SU(5) \subset SU(10)$.  These GUTs always enable one to absorb $e\,A = A'$.  We can only absorb $e$ into $A_\mu$ if every field couples with the same coefficient as in most GUTs.

\subsection{The Hidden-Spatial-Geometry School}
\label{Sec2.3}
The next school maps a hidden spatial geometry onto the gauge fields.  The gauge potential transforms inhomogeneously and makes unclear the physical nature of the theory.  In 1978, Goldstone and Jackiw \cite{Goldstone:1978he} made the electric field in an $SU(2)$ gauge theory diagonal, which made easier separating the gauge-invariant parts of gauge angles.  They wed these ideas to a 4-space `spinning top' analogy.

In 1994 Lunev \cite{Lunev:1994ty} formulated a tetrad-based mapping from $A^a_j$ to a tetrad variable.
In 1995 Freedman, Haagensen, Johnson, and Latorre also introduced tetrad variables $u^a_j$ as a replacement to $A^a_j$ \cite{Freedman:1993mu,Freedman:1994rg,Haagensen:1994sy,Haagensen:1995py}, where the index $a$ denotes the color index and $j$ denotes the spatial index. In our work we follow the notation of Haagensen and Johnson.  The $u^a_j$ variables serve as a mapping from the basis vectors that span an internal color space at a space-time point to the coordinate tangent vectors of a hidden spatial metric at that space-time point.  For this approach to work, the color-space dimension of the tetrad must be equal to the space-time dimension of a chosen slice.

Haagensen and Johnson used an $SU(2)$ gauge group, with structure constants $f^{abc} = \varepsilon^{abc}$ for the color index and $GL(3,\mathbb{R})$ for the spatial component.  They worked in the temporal gauge $A^a_0 = 0$ so they could map the three vector potentials $A^j_a$ to the three dimensions of the space slice.  The constraint imposed on $u_j^a$ was that the color index had to transform as a covariant vector under $SU(2)$ and the spatial index had to transform as $GL(3,\mathbb{R})$.  The condition that $u^a_j$ transform as a vector leads to the gluon `spin' operator constraint
\begin{equation}
\label{constraint}
\varepsilon^{ijk}(\partial_j u^a_k + \varepsilon^{abc}A^b_j u^c_k) = 0,
\end{equation}
which is similar to the spinning top analogy given in Jackiw and Goldstone.  The end result is that, for a given set of tetrads $u^a_j$, a unique vector potential $A^a_j$ can be found; however, the other direction is not unique.  For a given $A^a_j$ several $u^a_j$ exist.  Given a set of tetrad fields $u^a_j$, the $SO(3)$ vector potential is given by
\begin{equation}
A^a_j = \frac{(\varepsilon^{nmk}\partial_m u^b_k)(u^a_n u^b_j - \frac{1}{2} u^b_n u^a_j)}{{\rm{det}}\;u}.
\end{equation}

In using the constraint of Eq.~(\ref{constraint}), a hidden spatial metric was implicitly introduced.  The anti-symmetric tensor in Eq.~(\ref{constraint}) implies
\begin{equation}
\label{connect1}
\partial_j u^a_k + \varepsilon^{abc}A^b_j u^c_k = \Gamma ^s_{jk} u^a_s,
\end{equation}
where $\Gamma ^s_{jk}$ is a quantity symmetric in the indices $j,k$.
Notice that Eq.(\ref{connect1}) is the standard covariant derivative of a tetrad. It therefore implicitly defines the relationship between spin-connection $A_\mu$, the Levi-Civita-connection $\Gamma^s_{ij}$, and the tetrad $u^a_k$.

Standard manipulation shows that $\Gamma^s_{ij}$ is indeed the Christoffel symbols of the Levi-Civita-connection for a Riemannian manifold:
\begin{equation}
\Gamma ^i_{jk} = \frac{1}{2} g^{im}(\partial_j g_{mk} - \partial_k g_{jm}-\partial_m g_{jk}),
\end{equation} where $g_{ij} = u^a_i u^a_j$.
Thus, imposing Eq.~(\ref{constraint}) implicitly introduced a covariant derivative of a tetrad in Eq.(\ref{connect1}), and therefore a Riemannian geometry with a tetrad $u$ and a metric.

The matter fields are vectors in the tangent space of the manifold,
\begin{equation}
\vec{\phi} = \phi^i \vec{t}_i.
\end{equation}
Towards the end of the nineties Schiappa adapted these local gauge-invariant variables for supersymmetric gauge theory \cite{Schiappa:1997yh}.  An independent variation of this school was pursued by Slizovskiy and Niemi \cite{Niemi:2010mw}.

In summary, the variables $u^a_j$ map basis vectors that span the internal color space to coordinate tangent vectors of a hidden spatial metric.

\subsection{A Hidden Time Dependence to Electric Fields}

\label{SecHiddenTimeDependence}

At first it seems odd to suggest that a static electric field has a time dependence.
If we have a single non-accelerating charge, Coulomb's law produces a static electric field.
As nothing is moving, one would not expect any time dependence.
When we introduce gauge fields, we can either choose to describe a static electric field as the negative
gradient of a static voltage or as the time derivative of $\vec{A}$.
The two descriptions are related by a gauge transformation, but only one has an explicit time dependence.
Is this time dependence real or an artifact of poor choice of gauge?

The language of gauge theory has long suggested that the time dependence of an the electric field is fundamental but often hidden.
The electric field is given by the $0$-$i$ component of the field strength tensor which geometrically measures curvature of an internal space
after parallel transporting a wave function through a part spatial and part temporal space-time loop.
For this curvature to be non-zero, it seems that something must be changing with time.
The Lagrangian is typically written as the kinetic energy minus the potential energy.  In gauge theories the Lagrangian density ${\mathcal{L}}= \frac{1}{2} (E^2 - B^2)$ has the electric field play the role of kinetic energy.  This again suggests there may be a time-dependence to the electric field.

The explicit time dependence for electric fields in the Grassmannian school can be seen in the work Dubois-Violette and Georgelin \cite{Dubois-Violette:1979it}.  They expressed the field strength $F_{\mu\nu}$ in terms of the projection operators
 $P(x)^j_k = e^j_a(x) e^{\;a}_k(x)$ formed from the outer-product of the Grassmannian school's basis vectors.
Their expression
\begin{equation}
 e_a^{\,k} (F_{\mu\nu})^a_{\,b}e^b_{\,j} = ( P(x) [\partial_\mu P(x), \partial_\nu P(x)])^k_{\,j}
\end{equation}
shows that for $F_{0i}$ to be non-zero, then at a minimum
$\partial_0 P(x)$ must be non-zero.
This means if there is a non-zero electric field, then  the vector-space spanned by the
gauge fiber as seen in the Grassmannian school will be time-varying.

Is this time dependence an artifact of the Grassmannian representation?   What does it look like? The explicit time-dependence is not unambiguously present in the traditional field-strength description $F_{0i}$, nor in  the vector potential $A_\mu$, nor in the Kaluza-Klein representation, nor in the Hidden-spatial metric representation.  By providing the mapping between these different geometrical representation schools in the following sections, we hope to show that this time dependence should be taken more seriously.  In the subsequent examples, we'll be able to visualize a few special cases of this time-dependence in all three schools discussed in this paper.

\section{Mapping the Grassmannian School onto the Kaluza-Klein School}
\label{sec3}
We now wish to map the Grassmannian school of section \ref{Sec2.2} to the Kaluza-Klein school of section \ref{Sec2.1}.
 We begin with the Grassmannian school representation of an $SO(2)$ gauge theory.
We then construct an explicit isometric immersion into an $(4+N)$-dimensional Lorentzian space.
Finally, we calculate the induced $5$-dimensional metric.
This induced metric will be the Kaluza-Klein $5$-dimensional metric with the
the gauge field $A_\mu$ in the $\mu$ $5$ off-diagonal element of the metric
\begin{equation}
g_{\mu 5} = \vec{t}_\mu \cdot \vec{t}_5 \propto A_\mu \label{Eqgmu5}
\end{equation}
as is required in Kaluza-Klein theory.
The domain of the map is the Grassmannian schools representation given by $\vec{e}_a(x)$ where
 the vector potential is given by Eq.(\ref{A}).
 Narasimhan and Ramanan \cite{Narasimhan1961,Narasimhan63} and the additional references in section \ref{Sec2.2} showed that
 this rectangular matrix can be found for any vector potential $(A_\mu)^a_{\,b}$.
The final target or image of the map will be the $5$-dimensional Kaluza-Klein metric.
The generalization to non-abelian gauge fields is straight forward.

The first step of the map is that we insert the traditional space-time manifold and gauge fiber into an $SO(1,3+N)$ embedding with a Lorentzian signature $\eta={\rm{diag}}(-1,1,1, \ldots, 1)$.
The explicit embedding being considered is
\begin{equation}
\vec{X} = \left(t,x,y,z,R \vec{e}_{\,1} \cos \frac{x_5}{R} + R \vec{e}_{\,2} \sin \frac{x_5}{R} \right),
\label{EqKKEmbedding}
\end{equation}
where $\vec{e}_a$ is the rectangular $N \times n$-dimensional matrix from Grassmannian school explained in section \ref{Sec2.2}.
Because we are mapping an $SO(2)$ gauge theory to a Kaluza-Klein metric, the index $a$ will only run from $1$ to $2$.
The variable $x^5$ is the fifth Kaluza-Klein space-time coordinate which runs from $0$ to $2\,\pi\,R$ in our notation.
As is true for the Grassmannian school, the matrix $\vec{e}_a(x)$ depends only on the first four space-time coordinates $x^\mu$.
This inserts a ring in the embedding space.  The tangent vectors used in Eq.~(\ref{Eqgmu5}) are given by $\vec{t}_A = \partial_A \vec{X}$.  For the first 4 space-time coordinates the tangent vectors $\vec{t}_\mu$ are given by
\begin{equation}
t_\mu^k = \partial_\mu X^k = \delta^k_\mu + \Theta(k-5)\, R\, \left( \cos(\frac{x^5}{R}) \partial_\mu e_1^{k-4} + \sin(\frac{x^5}{R}) \partial_\mu e_2^{k-4}   \right),
\end{equation}
where the discrete Heaviside function $\Theta(x-a)$ is defined to be $1$ if $x \geq a$ and $0$ when $x<a$ and $k$ indexes the $4+N$ coordinates of the embedding space.
The tangent vector $\vec{t}_5$ is given by
\begin{eqnarray}
t^k_5 & = & \Theta(k-5)\, \partial_5\left( R\,e^{k-4}_1 \cos( \frac{x^5}{R}) + R\,e^{k-4}_2 \sin( \frac{x^5}{R}) \right) \nonumber \\
 & = &  \Theta(k-5)\,\left( -e^{k-4}_1 \sin( \frac{x^5}{R}) + e^{k-4}_2 \cos( \frac{x^5}{R})  \right).
\end{eqnarray}
The resulting five-dimensional space-time metric $\tilde{g}_{AB}$ for this embedding to first order in $R$ is
$ g_{\mu \nu} = t^k_\mu t^l_\nu \eta_{kl} = \eta_{\mu\nu}+ O(R^2)$,
$ g_{\mu 5} =  t^k_\mu t^l_5 \eta_{kl} = R A_\mu$, and $g_{55} = t^k_5 t^l_5 \eta_{kl} = 1$
where we have used Eq.~(\ref{A}) from the Grassmannian school applied to $SO(2)$ to relate $A_\mu = \vec{e}^{\;2} \cdot \partial_\mu \vec{e}_1 = -\vec{e}^{\;1} \cdot \partial_\mu \vec{e}_2$ and $\vec{e}_a = \vec{e}^{\;a}$.
The dilaton field $\Phi(x)$ follows if we allow the size of the curled up dimension to vary: $R \rightarrow R \Phi(x)$.
This $5$-dimensional metric is the target of this explicit map between these
previously defined geometrical representations of gauge theory.

A few general comments.
The geometry of the embedding, which reproduced the Kaluza-Klein metric, has a compactified ring at every point in space-time on the same plane spanned by the Grassmannian-school's basis vectors.
Visual examples will be shown in section \ref{Sec5}.

Although we have shown a general map to first order in $R$ between the Kaluza-Klein theory and the Grassmannian school, the Kaluza Klein school has a different representation of the wave function.
In the Grassmannian school there is one wave function at a space-time point and it is a vector on the tangent space spanned by the basis vectors $\vec{e}_a$.
In the Kaluza-Klein picture we see that the wave function is a function of each point in space-time including $x_5$.
It can vary circularly as we vary position of the fifth coordinate.  It is this feature that is responsible for the Kaluza-Klein tower of masses $m^2 = (n / R)^2$, where $n$ is again an integer.  The other schools lack such a mass tower.

Let us discuss the coordinate dependence and independence of the relationship between the Kaluza-Klein and Grassmannian models.  In the Kaluza-Klein school, the value of the $x^5$ coordinate is the same as the $\theta$ that delineates the angle on the ring inserted in the Grassmannian school.  In the immersion Eq.(\ref{EqKKEmbedding}), coordinate changes such as gauge transformations leave the surface formed by the immersion unchanged.
This does not mean that the surface  in Eq.(\ref{EqKKEmbedding}) is unique: there are many surfaces that lead to the same gauge field $A_\mu$.\footnote{ Some specific many-to-one mappings will be provided in section \ref{Sec5} in Eqs.(\ref{EqChargedSphereEmbeddingGeneral}) and (\ref{EqChargedSphereEmbeddingSpecific}).}
The many to one relationship does not mean that there is a coordinate dependence to the mapping.
Notice that all coordinate transformations leave the surface formed by immersion unchanged.
For example, see Figs.~3 and 5 of Ref.~\cite{Serna:2002ux}.  Fig.~3 of Ref.~\cite{Serna:2002ux} shows two different Grassmannian representations for the magnetic field of a solenoid. Both representations give the exact same gauge field, but they are not related by a gauge transformation.  Fig.~5 of Ref.~\cite{Serna:2002ux} shows the same magnetic field in two different gauges.  You can see the surface-like structure is unchanged by the change of gauge.

\section{Mapping the Hidden-Spatial-Geometry School onto the Grassmannian School }
\label{sec4}
Next we show how the hidden-spatial-geometry school of section \ref{Sec2.3} is mapped onto the Grassmannian school of section \ref{Sec2.2}.
The domain of the mapping is the tetrads $u^a_j$ of section \ref{Sec2.3}.  The target space will be the Grassmannian school representation.

The first step of the mapping is to use the Nash embedding theorem \cite{Nash56} to
define an immersion of the hidden spatial metric of section \ref{Sec2.2} into a larger-dimensional Euclidean space.  Nash guarantees that such an immersion exists for any metric given the dimension $N$ of the Euclidean space is sufficiently large.
Given this guaranteed immersion, the coordinate tangent vectors $\vec{t}_j$ of dimension $N \times n$ will reproduce the hidden-spatial-geometry metric

\begin{equation}
g_{jk} = u^a_j u^a_k = \vec{t}_j \cdot \vec{t}_k.
\end{equation}

Next, we identify the $N$-dimensional embedding-space dimensions that Nash guarantees exist with the trivial $N$-dimensional vector bundle used by Narasimhan and Ramanan in the Grassmannian representation.
The tetrads $u^a_j$ from the hidden-spatial-metric school will map the coordinate tangent vectors $\vec{t}_j$ to the orthonormal frame $\vec{e}_a$:
\begin{equation}
\label{tetrads}
\vec{e}_{a} = u^{\;i}_a \vec{t}_i,
\label{EqTetradGrassmanMap}
\end{equation}
or its dual,
\begin{equation}
\vec{e}^{\;a} = u^a_i \vec{t}^{\;i}.
\end{equation}
The tetrads $u^i_a$ may be obtained by using the Gram-Schmidt orthogonalization process on the coordinate tangent vectors of the hidden spatial metric.

The target space of the mapping is this orthonormal frame $\vec{e}^a_j$ which we will show is exactly the defining $N \times n$ rectangular matrix of the Grassmannian school representation.



Repeating the definitions from the literature presented in section \ref{Sec2.3}, we note that the color-space dimension of the tetrad must be equal to the dimension of the space-time slice under consideration.
For $SO(3)$ we need a three-dimensional slice of space-time to identify with the three color dimensions of the $SO(3)$ gauge fiber.
For the $SO(2)$ representation used in section \ref{Sec5}, we need a two-dimensional slice of space-time to identify with the two real dimensions of $SO(2)$.

We have proposed that Eq.(\ref{EqTetradGrassmanMap}) maps the hidden-spatial-metric school to the Grassmannian school.
To verify this claim, we will use Eq.(\ref{EqTetradGrassmanMap}) in the Grassmannian definition of the gauge field $A_\mu$ from Eq.(\ref{A}). We will check that it reproduces the defining relation in section \ref{Sec2.3}.
We express Eq.~(\ref{A}) not in terms of the gauge basis vectors but the coordinate tangent vectors $\vec{t}_i$ associated with the hidden spatial metric of the Grassmannian school via Eq.~(\ref{tetrads}), then Eq.~(\ref{A}) becomes
\begin{equation}
\label{substitution}
(u^a_i \vec{t}^{\;i}) \cdot \partial_j(u^k_b \vec{t}_k) = -i A_{j\;\;b} ^{\;\;a},
\end{equation}
\begin{equation}
\label{temp1}
u^a_{i} \delta^{i}_k \partial_j u^{k}_ b + i A_{j\;\;b} ^{\;\;a}  + u^{a}_i u^{k}_b \Gamma^{i}_{jk} = 0.
\end{equation}
Multiplying by $-u^b_l$ and using the identity $u^k_b \partial_j u^a_k + u^a_k \partial_j u^k_b = 0$ yields
\begin{equation}
\partial_j u^a_{k} - i A_{j\;\;b} ^{\;\;a}u^b_l \delta^l_k  - u^{a}_i \Gamma^{i}_{jk} = 0. \label{EqTetradGeneral}
\end{equation}
Now we specialize to $SU(2)$, where the form of the generators are\footnote{The distinction between lower and upper indices are dropped in the epsilon term for convenience (see Weinberg \cite{weinberg1996quantum}, chapter 15 appendix A).}
\begin{equation}
(T^c)^a_{\;\;b} = -i\varepsilon^{abc}.
\end{equation}
Thus Eq.~(\ref{substitution}) leads to
\begin{equation}
\label{SO3}
 \partial_j u^a_{k} + \varepsilon^{abc} A^b_j u^c_k   = u^{a}_i \Gamma^{i}_{jk},
 \label{EqHSGGrassmanianVersionTetradExpression}
\end{equation}
which is Eq.~(\ref{connect1}) of the hidden-spatial-metric school, but this time derived by the Grassmannian school's methods.  A similar relationship was independently observed by Refs.~\cite{Schuster:2003kt,2006JPhA...39.9187G} but without noting the generality of the relationships to the long research records of the two schools.

As for the wave function in the Grassmannian school, it is a vector on the gauge fiber in the internal space.  In the hidden-spatial-metric school, it is a vector on the tangent space.  As the gauge fiber is the same vector space in the two schools, then the wave functions are the same vector in these two schools.  This is in contrast to the Kaluza-Klein model, where the wave function is a scalar function of each point in the five-dimensional space-time.

Now let us discuss the coordinate dependence and independence of the relationship between the Grassmannian school and the hidden-spatial-metric school.  When formulated at MIT, there was no embedding space associated with the hidden-spatial-metric school.  Johnson, Haagensen, Schiappa, \emph{et.al.} highlighted that the metric formed by contracting over the color indicies in the tetrad $g_{ij}= u^a_i \,u^b_j \, \delta_{ab}$ was invariant under gauge transformations which only act on the internal color indices.  They also discussed the many-to-one relationship between metrics and gauge-fields.  Nash's \cite{Nash56} embedding theorem guarantees that we can represent the metric that represents the hidden-spatial-metric school as an isometric immersion into a trivial embedding space.
We observe that in the Grassmannian school, the tangent plane to the coordinates of a space-time point of the hidden-spatial metric, as viewed by the embedding, provide the element of the Grassmannian that corresponds to that space-time point.  Coordinate changes on the space-time slice do not change the surface.  Gauge-transformations do not change the metric nor the element of the Grassmannian that represents that point.
There are no special coordinates that enable the relationship between Eq.(\ref{connect1}), which was derived from symmetry principles without an embedding space, and Eq.(\ref{EqHSGGrassmanianVersionTetradExpression}), which was derived from a surface immersed in the embedding space guaranteed by Nash.  The mapping is general and does not depend on special coordinates.

\section{Examples in Electromagnetism}
\label{Sec5}
We now apply the geometric representations from the different schools to an abelian $U(1)$ gauge theory, namely ordinary electromagnetism.  We work with $SO(2)$ (which is isomorphic to $U(1)$) so that everything is real.  In $SO(2)$, there is only one generator; the gauge potential is
\begin{equation}
A_{j\;\;b}^{\;\;a}=  T^{a}_{\;\;b} A_j,
\end{equation}
where $T^a_{\;\;b} = -i\varepsilon^{ab}$.

Each electromagnetic field configuration has at least one (sometimes many) geometric representations in each school.  In order to demonstrate the hidden spatial geometry, we need to select a space-time slice of equal dimension to the dimension of the gauge fiber.  For $SO(2)$ we will need to select two-dimensional slices.  We will analyze two-dimensional slices of space-time denoted by $x^\mu (\sigma,\tau)$ and show each school's representation in this slice.  We use the pullback to map the four-dimensional field-strength tensor $F_{\mu\nu}$ and vector potential $A_\mu$ onto the 2-D slice of space-time using
\begin{equation}
\label{pullback}
F_{ij} = \frac{\partial x^\mu}{\partial x^i}\frac{\partial x^\nu}{\partial x^j} F_{\mu\nu},
\ \ \ \ A_{j} = \frac{\partial x^\mu}{\partial x^j} A_{\mu}.
\end{equation}

The $SO(2)$ analog of Eq.~(\ref{SO3}) is
\begin{equation}
\partial_j u^a_{k} + \varepsilon^{ab} A_j u^b_k  - u^{a}_i \Gamma^{i}_{jk} = 0.
\end{equation}
Solving this for $A_j$ gives
\begin{equation}
  A_j = \frac{-1}{2} \epsilon_{ac}\,g^{kl} u^{c}_l  (\partial_j u^a_k- \Gamma^i_{jk} u^a_i)  .
 \label{EqHSGMapToA}
\end{equation}

The Grassmannian and Kaluza-Klein school's equations are unaltered in specializing to $SO(2)$ examples.

We now proceed to illustrate the above connections for three elementary electromagnetic field configurations: a $y$-polarized plane wave, an electrically charged ring, and a spherical charge.  These examples were chosen for their familiarity and because their hidden spatial metrics correspond to a sphere, a paraboloid, and a funnel-shaped object respectively.

\subsection{The $Y$-Polarized Plane Wave}
Consider the four-potential for a $y$-polarized plane wave traveling in the $x$-direction
\begin{equation}
\label{PWA}
A_y = A_0 \cos(k(x-t)),
\end{equation}
where $A_0 = \frac{E_0}{k} = \frac{B_0}{k}$.  We take a $yt$ slice of space-time.  This is parameterized by $t(\sigma,\tau) = \tau, x(\sigma,\tau) = x_0, y(\sigma,\tau) = \sigma, $ and $ z(\sigma,\tau) = z_0$, where $x_0$ and $z_0$ are fixed coordinates.

Using the pullback  the $SO(2)$ vector potential for the plane wave is
\begin{equation}
\label{asphere}
 \;\; A_\sigma = A_0\cos(k(x_0-\tau)).
\end{equation}
%

The question now is: What two-dimensional shape from the hidden-spatial-metric school has this specific vector potential?  Using trial and error, we considered shapes until we found the ones whose tangent vectors led to Eq.~(\ref{asphere}).  The plane wave follows from a sphere parametrized as
\footnote{ \label{FootNoteX} We have reused the variable name $X$ to parametrize each immersion. This is not not same immersion as Eq.(\ref{EqKKEmbedding}) nor the same as in the other examples.}
\begin{equation}
  \label{sphere}
\vec{X} = \left(
              \begin{array}{c}
                \varrho \sin (k(x_0-\tau)) \cos (A_0\sigma) \\
                 \varrho \sin (k(x_0-\tau)) \sin (A_0\sigma) \\
                \varrho \cos(k(x_0-\tau)) \\
              \end{array}
            \right),
\end{equation}
where $\sigma=y$ and $\tau=t$, and $\varrho$ is a positive real value on which $A_i$ and $F_{ij}$ do not depend.

\begin{figure}
                 \centering
                 (a)\includegraphics[width=0.5\textwidth]{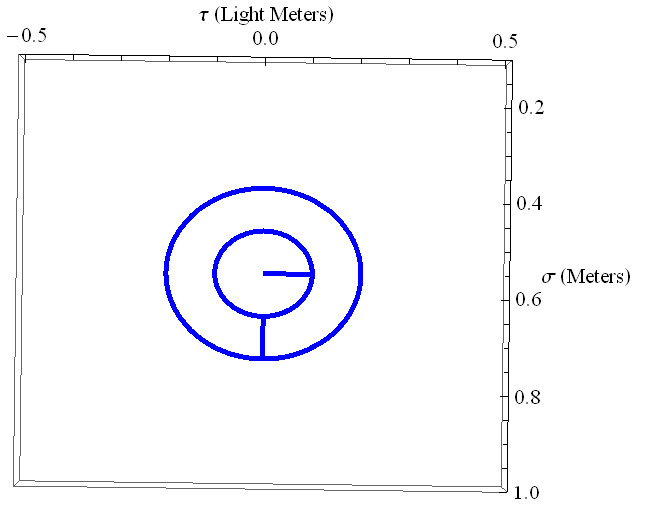}
                 \centering

                 (b)\includegraphics[width=0.8\textwidth]{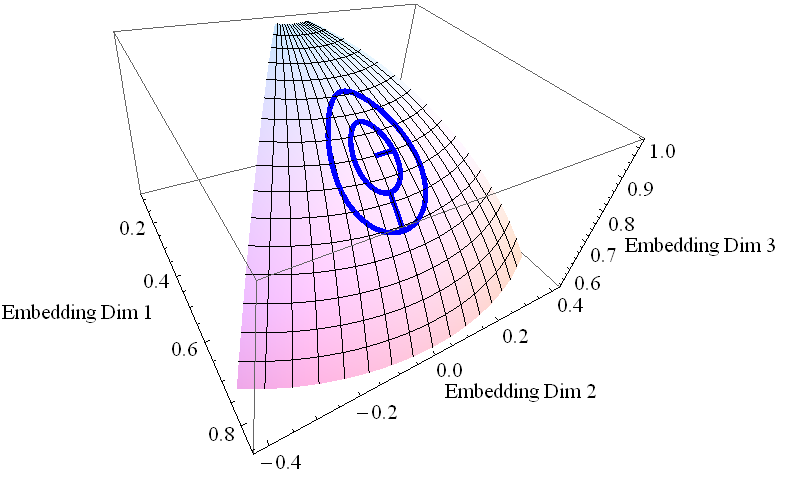}

                 \centering
                 (c)\includegraphics[width=0.65\textwidth]{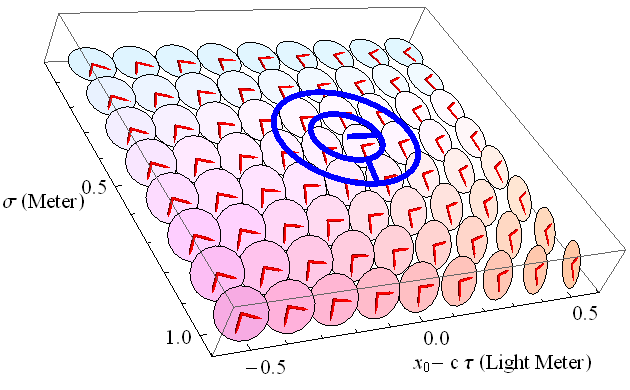}

                 (d)\includegraphics[width=0.8\textwidth]{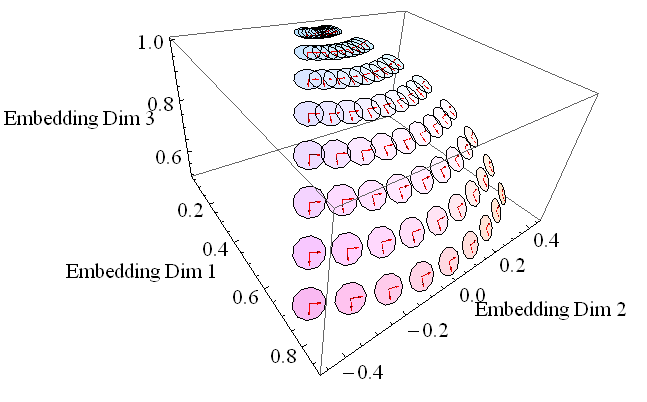}
                  \caption{\label{plainfigure}\label{PWshape}\label{planewave}\label{PWdisks}\label{PWdisksshape} The geometry of a $y$-polarized plane wave for a $yt$ slice, with $k=A_0=1\;{\rm{m^{-1}}},\varrho=1 \;{\rm{m}}$.
(a): Reference pattern which will be shown parameterized on the geometries of the three schools.
(b): A hidden-spatial-geometry representation of the $y$-polarized plane wave, $yt$ slice.
(c): A Grassmannian representation of the $y$-polarized plane wave, $yt$ slice.
(d): A hidden-spatial-geometry representation with the disks of the Grassmannian representation mapped to their corresponding location of the shape. }
\label{PW2}
\end{figure}

Fig.~\ref{plainfigure}a shows the domain of the variables $\sigma$ and $\tau$, which parameterize the $yt$ slice of space-time. This shape in Fig.~\ref{plainfigure}a helps map space-time points to the corresponding locations in the other figures. There are two lines, one in the direction of increasing $\sigma$ on the outer ring and one in the direction of increasing $\tau$ on the inner ring.  Fig.~\ref{PWshape}b shows the diagram as it appears parameterized on the surface of the hidden spatial geometry where we let $\varrho=1$ m, $k=1 \;{\rm{m}}^{-1},$ and $A_0=1 \;{\rm{m}}^{-1}$.  This corresponds to an average intensity beam of about $5 \times 10^{-17}\; {\rm{Watt/m^2}}$.
Here we see that increasing $\sigma$ corresponds to a line of longitude on the sphere, whereas increasing $\tau$ corresponds to a line of latitude.

Let us verify that the embedding from Eq.~(\ref{sphere}) produces Eq.~(\ref{asphere}).  The coordinate tangent vectors, $\vec{t}_\sigma$ and $\vec{t}_\tau$, are found by differentiating Eq.~(\ref{sphere}) by the respective coordinates $  \vec{t}_\tau = \partial_\tau \vec{X}$ and $  \vec{t}_\sigma = \partial_\sigma \vec{X} $.
The hidden-spatial-metric $g_{ij}$ is found by taking the dot products between each of the tangent vectors $g_{ij} = \vec{t}_i \cdot \vec{t}_j$.
The resulting line element is
\begin{equation}
ds^2 = k^2\varrho^2 d\sigma^2 + A_0^2\varrho^2 \sin^2(k(x_0-\tau)) d\tau^2.
\label{EqLineElementPlaneWave}
\end{equation}

This is the metric of the surface shown in Fig.~\ref{PWdisksshape}b.  Notice that each space-time points $(\sigma, \tau)$ correspond to points on the shape in Fig.~\ref{PWdisksshape}b.
The shape as parameterized is not the curvature of space-time, but a surface whose curvature represents the electric and magnetic fields of a plane wave.
If we were to perform a change of coordinates on $(\sigma,\tau) \rightarrow (\sigma', \tau')$ neither the shape nor the electric and magnetic fields would change.  This is because a point on surface of Fig.~\ref{PWdisksshape}b maps to a point on space-time. Coordinate re-parameterizations leave this mapping unchanged.  If instead one were to map the points on the surface to different space-time points, then the resulting electric and magnetic fields would be potentially very different.

Next we find the Grassmannian school representation. The lack of off-diagonal terms in the line-element of Eq.(\ref{EqLineElementPlaneWave}) means that the tangent vectors are orthogonal (such is generally not the case for coordinate tangent vectors, the sphere is kind enough to permit this simplicity).
The gauge basis vectors $\vec{e}_a$ on the gauge fibers are orthogonal and normalized.  While $\vec{t}_\sigma$ and $\vec{t}_\tau$ are orthogonal, they are not normalized. The tetrads $u^j_a$, which map $\vec{t}_j$ to $\vec{e}_a$, are $u^\sigma_1 = \frac{1}{|\vec{t}_\sigma|}=\frac{1}{k\varrho}$, $u^\sigma_2 = 0$, $u^\tau_1=0$, and $u^\tau_2 = \frac{1}{|\vec{t}_\tau|}=\frac{1}{A_0\varrho\sin(k\tau)}$.

Normalizing the tangent vectors gives the Grassmannian's basis vectors
\begin{equation}
\vec{e}_1 =  \left(
              \begin{array}{c}
                - \sin (A_0\sigma) \\
                  \cos (A_0\sigma) \\
                         0 \\
              \end{array}
            \right), \ \ \ \
\vec{e}_2 = \left(
              \begin{array}{c}
                - \cos (k(x_0-\tau)) \cos (A_0\sigma) \\
                 - \cos (k(x_0-\tau)) \sin (A_0\sigma) \\
                 \sin(k(x_0-\tau)) \\
              \end{array}
              \right).
              \label{EqPlaneWaveGrassmannian}
\end{equation}

Now we map the Grassmannian representation back to the gauge field representation.  The only nonzero value of $A_j$ is
\begin{equation}
     (A_\sigma)^a_{\ b} = i \vec{e}^{\;a} \cdot \partial_\sigma \vec{e}_b = -i A_0 \cos(k(x_0-\tau))  \varepsilon^{ab}
\end{equation}
which shows that this parameterization of the sphere leads to the geometry of the $yt$ slice of the $y$-linearly-polarized plane wave.
Likewise if we calculate $A_j$ from the hidden-spatial-geometry school via the tetrads $u^a_j$ and Eq.(\ref{EqHSGMapToA}), we also find the plane wave.

Eq.(\ref{EqPlaneWaveGrassmannian}) is visually displayed in Fig.~\ref{PWdisksshape}c. We can see that each space-time point $(\sigma,\tau)$ has the same tangent plane as the corresponding space-time point in Fig.~\ref{PWdisksshape}b.
As we move toward smaller $\sigma$ on Fig.~\ref{PWdisksshape}c, we see the tangent planes approaching a common plane which maps to the north pole of Fig.~\ref{PWdisksshape}b.
These vectors $\vec{e}_1$ and $\vec{e}_2$ are visualized in Fig.~\ref{PWdisks}c as the red basis vectors which span the disks.  The figure is to be interpreted as in Fig.~\ref{Standardpic}, but without the bubbles.  The reference pattern from Fig.~\ref{plainfigure}a is again shown to help visualize the directions of $\sigma$ and $\tau$ in both spaces.
A rotation of the red basis vectors on the disks corresponds to a gauge transformation.

Next Fig.~\ref{PWdisksshape}d shows the disks from the Grassmannian-school representation  rearranged into the shape associated with  the hidden-spatial-metric school.  The reference pattern is removed, but one can see the tangent plane associated with each space-time point mapped across all three Figs.~\ref{PWdisksshape} b, c, and d.

\begin{figure}[t!]
         \centering
           \includegraphics[width=0.8\textwidth]{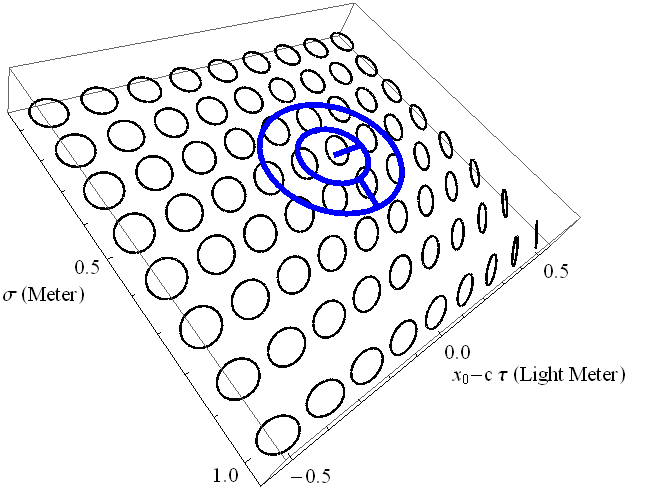}
           \caption{A Kaluza-Klein representation of the $y$-polarized plane wave from a $yt$ space-time slice.}
         \label{PWkaluza}
\end{figure}

From the Kaluza-Klein picture, we have at each point in space-time a curled up fifth dimension.  This is represented by a ring in the Grassmannian's space on the same tangent plane.  The ring's embedding is parameterized by $x^5$ as
\begin{equation}
\vec{r}(x^5) = R\,\vec{e}_1 \cos(\frac{x^5}{R}) + R\,\vec{e}_2 \sin(\frac{x^5}{R}),
\end{equation}
where $\vec{r}$ is a three-dimensional vector in a trivial Grassmannian's embedding space over space-time.  Fig.~\ref{PWkaluza} shows the representation of the Kaluza-Klein school, where the rings are shown at several space-time points for our given slice.  We have changed the scale of $R$ to better see the ring.  Thus, Figs.~\ref{planewave}b, \ref{PW2}c, and \ref{PWkaluza} are the three school's geometrical representations of the plane wave; all share a common set of tangent planes as represented in the embedding.

A shape by-itself does not represent an electric or magnetic field configuration.  The points on the shape must be mapped to space-time points.  If we change  the space-time point to which a point on the shape maps, then it manifests as a different gauge field. For example if we map a shape to an $xy$ slice of space, it will manifest as a $z$-directed magnetic field.  If we map a shape to an $xt$ slice of space-time it will manifest as an $x$-directed electric field.

As an example consider the $A_\mu$ of Eq.~(\ref{PWA}). Now we want to understand the magnetic field.  We consider the $xy$ slice of space-time: $t(\sigma,\tau) = t_0, x(\sigma,\tau) = \sigma, y(\sigma,\tau) = \tau, $ and $ z(\sigma,\tau) = z_0$
where $t_0$ and $z_0$  are fixed values.  The pullback of the vector potential on the $xy$ slice  is
\begin{equation}
 \;\; A_\sigma = \frac{B_0}{k} \cos(k(\sigma-t_0)).
\end{equation}
The shape corresponding to this slice is also a sphere $^{\ref{FootNoteX}}$:
\begin{equation}
  \vec{X} = \left(
              \begin{array}{c}
                \varrho \sin (k(\sigma-t_0)) \cos (A_0\tau) \\
                 \varrho \sin (k(\sigma-t_0)) \sin (A_0\tau) \\
                \varrho \cos(k(\sigma-t_0)) \\
              \end{array}
            \right).
            \end{equation}
To the best of our understanding, the similarity to Eq.~(\ref{sphere}) is not general.


\subsection{The Electrically Charged Ring}
\begin{figure}
         \centering
                 \centering
                (a) \includegraphics[width=0.85\textwidth]{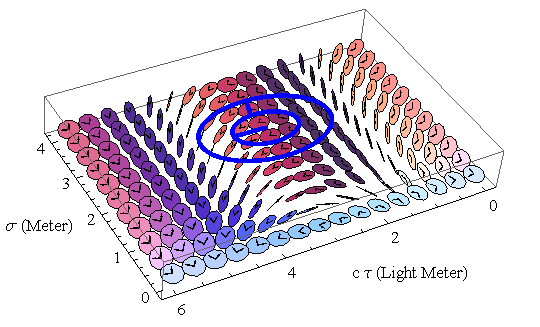}

                (b) \includegraphics[width=0.8\textwidth]{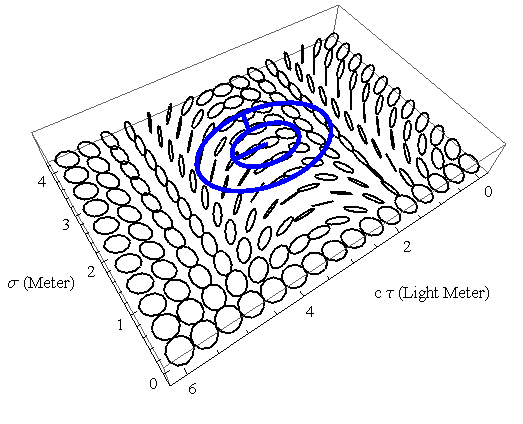}

                 \centering
                 (c) \includegraphics[width=0.95\textwidth]{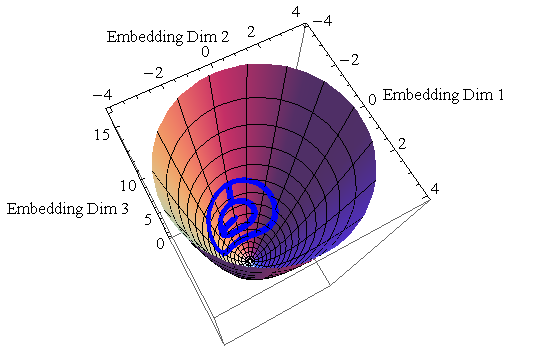}

\caption{ The geometry of an electrically charged ring, along the $z$ axis only, for $Q= 2 \pi$ and $b=\frac{1}{2}$ m. In SI units this is a charge of 1.24 $\mu$C.
(a): A Grassmannian representation showing the $\mathbb{R}^2$ subspaces along points of space-time.
(b): A Kaluza-Klein representation showing the gauge subspaces at each space-time point for an electrically charged ring.
(c): A hidden-spatial-geometry representation of the electrically charged ring.
\label{Paraboloiddisksshape}
\label{paraboloid}
\label{Paraboloiddisks}
\label{Paraboloidkaluza}
\label{ParaboloidShape}
}
\end{figure}
Next consider the electric field due to a ring of charge $-Q$ and radius $b$ centered at the origin on the $xy$ plane:
\begin{equation}
\label{ring}
E_z = \frac{-Qz}{4\pi(b^2 + z^2)^{\frac{3}{2}}}. \label{EqEzRing}
\end{equation}
The associated vector potential pulled-back onto a $zt$ slice ($z=\sigma$, $t=\tau$, $x=0$, $y=0$) is given by
\begin{equation}
A_\tau=\frac{Q}{4\pi} \frac{1}{\sqrt{b^2+\sigma^2}}. \label{EqVoltageRing}
\end{equation}
We found the hidden-spatial-geometry shape to be a paraboloid parametrized as $^{\ref{FootNoteX}}$:
\begin{equation}
\label{Xring}
               \vec{X}=\left(
                         \begin{array}{c}
                           \frac{\sigma}{2b} \cos(\frac{Q}{4 \pi b}\tau) \\
                            \frac{\sigma}{2b} \sin(\frac{Q}{4 \pi b}\tau)\\
                          ( \frac{\sigma}{2b})^2 \\
                         \end{array}
                       \right). \label{EqParabolidChargeRing}
\end{equation}
After calculating the tangent vectors $\vec{t}_j$ and the tetrads $u^a_j$ that create the basis vectors $\vec{e}_a$, we calculate the associated vector potential $A_\tau$ to verify that it gives the same $z$-directed electric field of the negatively charged ring.

Fig.~\ref{Paraboloiddisks}a shows the charged ring from the Grassmannian school with the choice of $b=\frac{1}{2}$ m and $Q = 2 \pi$.  The associated electric-field pictured corresponds to a 1.24 $\mu$C ring when converted to SI units.
Fig.~\ref{Paraboloidkaluza}b shows the gauge subspace from the Kaluza-Klein school, and Fig.~\ref{ParaboloidShape}c shows the hidden-spatial-metric picture.

Now let us carefully study these figures.
The electric field in Eq.(\ref{EqEzRing}) is constant in time.
The corresponding gauge field (voltage) shown in Eq.(\ref{EqVoltageRing}) is also independent of time.
If we had chosen a different gauge, \emph{e.g.} the temporal gauge with $A_0=0$, then there would be a linear time dependence.  Is this time-independence a gauge artifact then, or is it part of the actual physics about our charged ring? By using Fig.~\ref{ParaboloidShape} which shows a \emph{gauge-invariant} representation of the electric field, we can uncover the underling gauge-invariant time dependence common to all three representations.

In Fig.~\ref{ParaboloidShape}a one sees the tangent planes repeating a pattern as the coordinate $c \tau$ advances. Notice that this time dependency cannot be eliminated by a gauge transformation or clever coordinate choice. This corresponds to the periodicity of the cosine function with respect to $\tau$ in Eq.(\ref{EqParabolidChargeRing}).
The embedding space is a fixed reference which enables one to see the gauge-invariant phenomena.

From the Kaluza-Klein picture in Fig.~\ref{Paraboloiddisks}b, we also see a repetition in the disk arrangements in the $\tau$ direction.
This again follows from  to the periodicity of the cosine function with respect to $\tau$ in Eq.(\ref{EqParabolidChargeRing}).   Notice that at the metric-level, the terms which represent the gauge field $g_{5t} \propto\, R \,A_0$ do not depend on the time coordinate $\tau$.
When we view the Kaluza-Klein surface from an embedding, we can see that the off-diagonal terms follow from the time dependency of the orientation of the ring parametrized by $x^5$ as viewed from the embedding.

In the hidden-spatial-geometry school shown in Fig.~\ref{Paraboloiddisksshape}c,
we see that time dependence corresponds to an oscillation around a paraboloid shape.
The direction of increasing $\sigma$ is along the vertical dimension of the paraboloid.
As $\sigma$ increases the shape becomes flatter which corresponds to distances farther from the ring (with a weaker electric field along the axis).
The $\tau$ direction is along the circular dimension of the paraboloid.
Our position on the paraboloid changes with time showing the hidden-time dependence from another vantage.

Note that the charged ring has an explicit time dependence in all three gauge-invariant geometric representations, as shown by $t=\tau$ in Eq.~(\ref{Xring}).  The time dependence disappears when we project to the scalar potential and the electric field.  Although the charged ring gives a static electric field, the geometrical representations makes clear there is a hidden gauge-invariant time dependence.


\subsection{The Spherically Charged Shell}

\begin{figure}
         \centering
                (a) \includegraphics[width=0.65\textwidth]{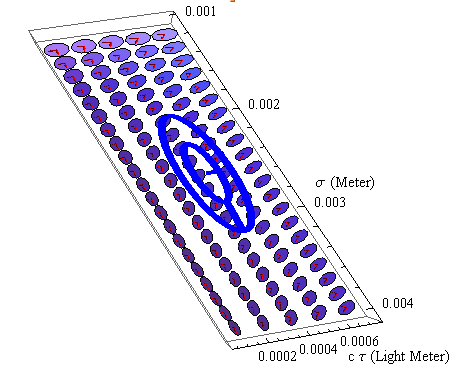}

                (b) \includegraphics[width=0.6\textwidth]{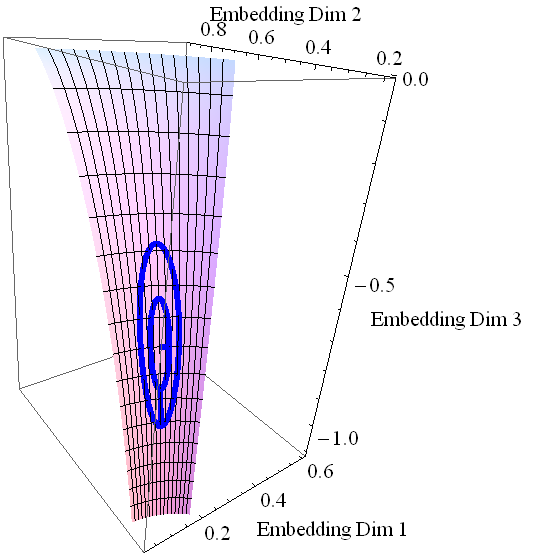}

                 \centering

                (c) \includegraphics[width=0.6\textwidth]{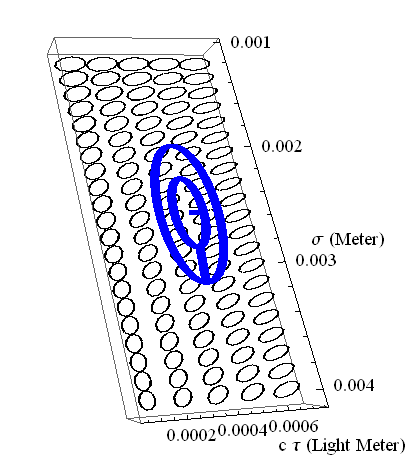}

       \caption{\label{Spheredisks} \label{Sphereshape}\label{Spherediskshape}\label{Spherekaluza}The geometry of a spherical charge for $q = 4 \pi$ and $\omega = 1\, {\rm{nm}}^{-1}$.
       (a): A Grassmannian school representation of the charged sphere.
       (b): A hidden-spatial-geometry representation of the charged sphere.
       (c): A Kaluza Klein representation which shows rings at each space-time point for a spherical charge.}
\label{Sphere}
\end{figure}
For a bounded scalar potential $A_t = \Phi(\vec{x})$ of a static electric field, a funnel-shaped surface can be found for a given two-dimensional slice of space-time.  In this case, $A_\mu = \left(
                                                             \begin{array}{cccc}
                                                               \Phi, & 0, & 0, & 0 \\
                                                             \end{array}
                                                           \right)
$, and $A_\tau =\frac{\partial t}{\partial \tau} A_t $ is a function of $\sigma$ only.  The first derivative of $A_0$ must be strictly negative, and $0 < A_\tau \leq \omega$.  If we use the shape $^{\ref{FootNoteX}}$
\begin{equation}
\label{gen}
 \vec{X}=\left(
                         \begin{array}{c}
                           \frac{A_\tau (\sigma)}{\omega} \sin(\omega\,\tau) \\
                            \frac{A_\tau (\sigma)}{\omega} \cos(\omega\,\tau)\\
                           \sqrt{1-(\frac{A_\tau (\sigma)}{\omega})^2} + \ln(\frac{\frac{A_\tau (\sigma)}{\omega}}{1+\sqrt{1-(\frac{A_\tau (\sigma)}{\omega})^2}})\\
                         \end{array}
                       \right)
 \label{EqChargedSphereEmbeddingGeneral}
\end{equation}
with a $t(\tau) = \tau, x^i = x^i(\sigma)$ slice of space.  The variable $\omega$ represents a continuous class of geometries which give rise to a single $A_\tau$.  Notice that $\omega$ must be nonzero and larger than the maximum value of $A_\mu$ in the given domain.

Consider a spherical shell of charge $q$ with radius smaller than $q/ 4 \pi \omega$, and let $x(\sigma) = \sigma, y = 0, z = 0$.  The only nonzero component of the field tensor, when looking strictly along the $x$-axis, is $F_{tx}=E_x = \frac{q}{4\pi x^2}$.  The pullback gives $F_{\sigma \tau} = \frac{q}{4\pi \sigma^2}$ and $A_\tau = \frac{q}{4\pi \sigma}.$  Using Eq.~(\ref{gen}), we find that the surface that is associated with the charged sphere, looking along the $x$-axis, is $^{\ref{FootNoteX}}$
\begin{equation}
 \vec{X}=\left(
                         \begin{array}{c}
                           \frac{q}{4 \omega \pi \sigma} \sin(\omega\,\tau) \\
                            \frac{q}{4 \omega \pi \sigma} \cos(\omega\,\tau)\\
                           \sqrt{1-(\frac{q}{4 \omega\pi \sigma})^2} + \ln(\frac{\frac{q}{4 \omega\pi \sigma}}{1+\sqrt{1-(\frac{q}{4 \omega\pi \sigma})^2}})\\
                         \end{array}
                       \right).
 \label{EqChargedSphereEmbeddingSpecific}
\end{equation}
Letting $q = 4 \pi$  and $\omega = 1\, {\rm{nm}}^{-1}$, Fig.~\ref{Spheredisks}a shows the spherical charge from the Grassmannian school,  Fig.~\ref{Sphereshape} b is the hidden-spatial-metric picture,
and Fig.~\ref{Spherekaluza}c shows it from the Kaluza-Klein picture.  In SI units this corresponds to the field of a $2.2 \times 10^{-17}$ C charge where $\sigma > 1$ nm.  From the reference Fig.~\ref{Spheredisks}a, we see that increasing $\sigma$ is down toward the tip of the funnel and increasing $\tau$ is on the circular dimension.  This makes geometrical sense, as $\sigma$ increases, we move towards the narrow throat of the funnel, and the shape gets more cylinder-like.  Given that we have potential $A_\tau = \frac{q}{4\pi \sigma},$ as we move farther from the sphere, the weaker the field becomes, leading to a less curved surface.

Finally, note that the value of $\omega\ [> {\rm{max}}(A_\tau)]$ is arbitrary in this case.
The time dependence of $\omega$, which is clearly present in all three gauge-invariant geometrical representations, vanishes in the scalar potential and electric field.  Again, the geometric relationships make it clear that there is a hidden gauge-invariant time dependence in the electric field of the charged sphere.

Let us study Fig.~\ref{Sphere} more carefully.  We are again showing a time-independent electric field, but we see time dependence when we dig down to the surfaces that underlie the 1-form connection.  The figures show a cutout in time.  If we had continued the figures towards larger $\tau$, you would again see a periodic pattern for all three schools.
In the Grassmannian representation shown in Fig.~\ref{Sphere}a, let us look at $\sigma=0.004$ as we vary $\tau$.  If we continued $\tau$ towards larger values, we would see the disks complete a complete cycle and repeat. A change of gauge will change the choice of red basis vectors that span the disks, but the disks (the actual element of the Grassmannian) remains unchanged.
Fig.~\ref{Sphere}b shows the hidden-spatial-geometry where we see the cutout of a funnel shape explicitly.  The $\sin \omega \tau$ and $\cos \omega \tau$ in Eq.(\ref{EqChargedSphereEmbeddingSpecific}) show that if we continued to plot points of larger $\tau$, we would fill out the funnel and begin to repeat.
Fig.~\ref{Sphere}c show the rings that live on the same tangent plane as the Grassmannian school.
All three figures show a surface-like geometrical representation that gives rise to the 1-form connection.  All three figures use the embedding space so that the geometrical objects (shape and disks) are independent of the gauge choice.  In all three representations, we can see the time dependence that is absent (or ambiguous) in the 1-form connection.  This does not mean that the figures here are unique.  The freedom to choose $\omega$ is an example of the many-to-one map that is associated with this phenomena.

\section{Discussion and Conclusion}

The field of gauge theory geometry is vast. Fig.~\ref{FigMapGaugeTheory} shows the curvature 2-form electric and magnetic fields as the layer with which most physicists are familiar.
By digging down to find the connection 1-form that gives rise to the curvature, physicists discovered the Aharonov-Bohm effect. We wish to dig one layer deeper.
We have grouped  into three schools the past efforts to find a surface-like layer that would give rise to
the connection 1-form.
Our paper shows the dotted-red line connections between these past efforts.

We have shown how these three representations of gauge theory that isolate gauge-invariant surface-like structures are related geometrically without appealing to their common gauge-field image space.
For the Kaluza-Klein school every point in space-time has a bundled-up fifth dimension.  With  an immersion given by Eq.(\ref{EqKKEmbedding}) that inserts a ring on the Grassmannian school's tangent-planes, we can recover the Kaluza-Klein metric and visualize this 5th dimension.  The Grassmannian school uses vector bundles to describe gauge fields.  We can visualize the subspace represented by these disks at each point in space-time using an embedding space inside at each space-time point.  Finally, by combining the embedding with a shape unearthed in the hidden-spatial-metric school, we can associate a spatial geometry with a gauge field configuration.

The similarities are deep.  All three schools share a common gauge-invariant tangent plane.  The wave function and projection operator are invariants of the gauge transformations.  This is because gauge transformations correspond to a rotation of the basis vectors that leave the tangent-plane unchanged.  This tangent-plane is the same plane in  both the Grassmannian and hidden-spatial-metric schools. Gauge transformations in Kaluza-Klein theory change the $0$ point of the $x^5$ coordinate on the ring defined in Eq.(\ref{EqKKEmbedding}).  The tetrads in the hidden-spatial metric school change the coordinate basis vectors of the hidden spatial geometry to orthonormal basis vectors on the gauge fiber. The Kaluza-Klein ring lies along the gauge fiber.  There are also similarities related to charge: in the Grassmannian and hidden-spatial-metric schools a positive or negative charge corresponds to a clockwise or counterclockwise rotation of a matter field vector on the gauge fiber, and in the Kaluza-Klein school a positive or negative charge corresponds to opposite movement along the ring (which lies along the gauge fiber).  The opposite charge corresponds to reverse rotation in all three schools.

Each of the three schools considered in this paper uses a gauge-invariant surface-like geometrical structure to induce the gauge field. Normally we must work very hard to isolate what is gauge-invariant.
For example, Killing-vector methods identify symmetries that can help reveal gauge-invariant physical results like conservation laws. We in contrast are digging down to reach the gauge-invariant surfaces which induce the gauge-dependent gauge field.
No Killing-vector-like approach is needed. In each of the three schools, one might have thought the resulting metric or Grassmannian representation was a special mathematical trick, and didn't say something very profound.
However we have shown how the surface-like structures underlying each school are related to each other.  This suggests that perhaps results regarding this surface-like foundation have some meaning.
As of now, they are just representations so no new physics should be present.
However, our new understanding of the mappings between the representations may help us ``better guess" new physical laws \cite{feynman1994character}.


In section \ref{SecHiddenTimeDependence} we showed how previously
published work indicated a time dependence for electric fields when represented in the Grassmannian school.
 The mappings we provide between the three
 schools show that this gauge-invariant time dependence exists for all three schools.
 This agreement suggests the hidden time dependence should be taken more seriously and may be a new feature at the foundation of physics.

In section \ref{Sec5}, we showed examples of what this time dependence `looks' like
in static electric fields.
In the charged ring and the charged sphere, we show that even static electric fields have a hidden, gauge-invariant time dependence  in the surface-like structures underling the gauge-field.
In both cases, it is a time-harmonic repeating wobble in the underlying surface.
Is this time dependence physical or an artifact of the geometric representation? The surface-like structure is common to all three schools.
The surface-like structure does not change with a gauge transformation. The time-dependence in the surface-like structures are therefore not an artifact of the parametrization or the coordinate system choice.


If we take the generalized time-dependent gauge-invariant features here seriously, there are several questions to pursue in future research.
 What is the underlying surface that is moving in the case of gauge theories.
What is the time-dependent source of the electric field's time dependence?
The wave function has a time dependent phase.  It would causally make sense if the wave function's time dependence was the source for the electric field's time dependence, but these two time dependencies are currently uncorrelated.
 We suspect that resolving this tension will force a deeper form of Guass's Law.
Are there any new observable consequences to this deeper Gauss's Law?  A deeper form of Gauss's law may also help address the mysteries of the mass gap in Yang-Mills theory.

Another area of development lies within the study of instanton and other less-well-known semi-classical solutions. Many of the papers previously published in each school were geared towards identifying and studying instanton solutions.
By using the mappings we identify, checking and categorizing instanton solutions may become easier, and it would be another tool for finding other semi-classical solutions that would need to be included in the path integral quantization.\footnote{The authors would like to thank Ricardo Schiappa for highlighting these research directions.}


\begin{acknowledgements}
The authors would like to thank Laura Serna, Kevin Cahill, Richard Cook, Matt Robinson, Christian Wohlwend, Ricardo Schiappa, and Yang-Hui He for helpful comments after reviewing the manuscript.  We would also like to thank the reviewers for helpful contributions increasing the quality of the final paper.
The views expressed in this paper are those of the authors and do not reflect the official policy or position of the United States Air Force, Department of Defense, or the US Government.  DISTRIBUTION A: Approved for public release. Distribution unlimited.
\end{acknowledgements}

\bibliographystyle{spmpsci}      

\newcommand{\noopsort}[1]{} \newcommand{\printfirst}[2]{#1}
  \newcommand{\singleletter}[1]{#1} \newcommand{\switchargs}[2]{#2#1}

\appendix

\section{Appendix: Variable definitions reference}

\label{AppendixVariableDefs}

\begin{tabular}{|c|p{8cm}|}
\hline
 $\vec{e}_a$  or $e_a^j$ & The basis vector for the Grassmannian school.  The $a$ coordinate is an internal `color' index.  If there is a Latin index like $j$, it refers to the embedding space. Forms a rectangular matrix.\\ \hline
 $\vec{t}_\mu$ or $t_\mu^j$ &  Coordinate tangent vector for the coordinate $x^\mu$. Used in defining a metric. If there is a  Latin index like $j$, it refers to the embedding-space dimension. Forms a rectangular matrix. \\ \hline
 $u^a_j$ & The tetrad of the hidden-spatial-geometry school.  Notice the $a$ index specifies the `frame' in color space and $j$ is a `frame' in a slice of space-time. This maps the color index $a$ to the space-time coordinate tangent vector $j$ of a spatial metric which represents the gauge field and corresponding electric and magnetic fields. Must be a square matrix. \\ \hline
 $\vec{X}$ or $X^j$ & Is the generic vector used to denote an explicit isometric embedding which will be used to induce a metric.  The Latin index $j$ refers to the embedding space.  \\ \hline
 $\phi^a$ &  Coefficients of the basis element $\vec{e}_a$ which specify a vector in color-space.  $\phi^a$ changes with a gauge transformation but the vector $ \vec{\phi} = \phi^a \vec{e}_a = \phi'^{\,b} \vec{e}'_b$ is gauge invariant. \\ \hline
 $ a, b, c$ & Lower-case Latin letters near the beginning of the alphabet will be gauge-theory color indices \\ \hline
 $\mu, \nu,...$ & Greek letters will be space-time coordinates\\ \hline
 $A,B,...$ & Upper-case Latin letters  will be used for Kaluza-Klein metric indices.  Kaluza-Klein index values 0 through 3 are the usual space-time coordinates $t,x,y,z$ and the index value 5 is the fifth dimension coordinate $x^5$, which is used to parameterize the tiny compact dimension. \\ \hline
 $i,j,...$ & Variables corresponding to subspaces of space-time and the embedding dimensions, where context will keep them distinct. \\ \hline
\end{tabular}

\end{document}